\documentclass[epj]{svjour}

\newcommand{\Ref}[1]{Ref.~\cite{#1}}
\newcommand{\Fig}[1]{Fig.~\ref{fig:#1}}
\newcommand{\eg}{\emph{e.g.}}
\newcommand{\ie}{\emph{i.e.}}
\usepackage{graphicx}
\usepackage{fancyhdr}

\def\mytitle{My title} 
\def\myauthors{My name}  
\def\mytype{My type of session}
\def\mysession{My session}

\def\mytitle{Neutrinos from WIMP annihilations} 
\def\myauthors{Mattias Blennow}    
\def\mytype{Contributed Talk}    
\def\mysession{Cosmology and Astrophysics}

\begin{document}
\title{Neutrinos from WIMP annihilations}
\author{Mattias Blennow
\thanks{\emph{Email:} emb@kth.se}
\thanks{Based on work done in collaboration with Joakim Edsj\"o and Tommy Ohlsson \cite{Blennow:2007tw}.}%
}                     
\institute{
Department of Theoretical Physics \\
School of Engineering Sciences \\
Royal Institute of Technology (KTH) \\
AlbaNova University Center \\
Roslagstullsbacken 21 \\
106 91 Stockholm \\
Sweden
}
%
\date{}
\abstract{
We make an improved analysis on the flow of neutrinos originating from
WIMP annihilations inside the Sun and the Earth. We treat both
neutrino interaction and oscillation effects in a consistent
framework. Our numerical simulations are performed in an event based
setting, which is useful for both theoretical studies and for creating
neutrino telescope Monte Carlos. We find that the flow of muon-type
neutrinos is enhanced or suppressed depending on the dominant WIMP
annihilation channel.
} 
\maketitle
\section{Introduction}
\label{sec:intro}

Weakly Interacting Massive Particles (WIMPs) are viable candidates for accounting for the dark matter (DM) in the Universe. While there are several experiments trying to directly detect or produce WIMPs, we will here focus on the indirect detection of WIMPs through the neutrino flux arising from WIMP annihilations in the Sun, which could be within the discovery reach of future neutrino telescopes such as IceCube. In general, WIMPs from the galactic DM halo will occasionally scatter on the solar constituents and become gravitationally bound to the solar system. Eventually, these WIMPs will loose increasing amounts of energy as they scatter repeatedly and sink to the solar core. Consequently, there will be an increased WIMP density in the solar core, leading to an increase of the WIMP-WIMP annihilation rate. Although neutrinos are not the only products of such annihilations, only neutrinos have cross-sections which are small enough for them to escape the Sun.

In the following, we will show the results of treating the neutrino propagation from creation in the Sun to interaction in the detector in a full three-flavor Monte Carlo. In particular, we focus on the effects of including neutrino oscillations and their impact on the final neutrino yield. For illustration, we will discuss the yields resulting from WIMPs with a mass of $m_\chi = 250$~GeV annihilating into $\tau^+\tau^-$ as this annihilation channel provides us with an initial tau neutrino yield which is significantly higher than the electron and muon neutrino yields (and thus, making the neutrino oscillation effects more pronounced). The resulting neutrino yields will be presented at four points; at the center of the Sun, at the solar surface, at 1~AU, and time-averaged at the detector. The full results of the Monte Carlo simulations performed in \Ref{Blennow:2007tw} are available at \Ref{homepage}.

\section{Neutrino production}

\begin{figure}
\includegraphics[width=0.45\textwidth,angle=0]{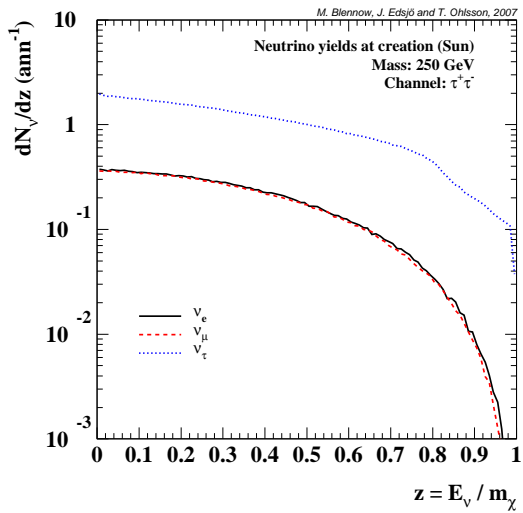}
\caption{The neutrino yields as a function of $z = E_\nu/m_\chi$ at creation in the center of the Sun for annihilation of 250~GeV WIMPs into $\tau^+\tau^-$. From \Ref{Blennow:2007tw}.}
\label{fig:production}
\end{figure}
In \Fig{production}, we show the neutrino yields at the center of the Sun. As can be observed in this figure, the tau neutrino yield is almost an order of magnitude larger than the yields of the other neutrino flavors at all neutrino energies. Note that the electron and muon neutrino yields will always coincide at the production point, but that the difference in these fluxes compared with the tau neutrino flux depends on the annihilation channel (\eg, the annihilation of WIMPs into $b\bar{b}$ results in a deficit of tau neutrinos).

\section{Neutrino interactions and oscillations}

In order for our results to make sense, we need to treat neutrino interactions and oscillations in a consistent framework. Since our Monte Carlo is event based, we follow each neutrino flavor state individually as it propagates. The neutrino oscillations are treated using the full three-flavor Hamiltonian including the MSW interaction term. During the evolution of states from inside the Sun, we approximate the solar electron number density by a number of layers with constant electron number density (each layer has a width of 0.3~\% of the solar radius). 
Using this approximation, we can compute the evolution matrix $S$ as the product of the evolution matrices for each layer as
\begin{equation}
S = S_n S_{n-1} \ldots S_2 S_1,
\end{equation}
where $S_i$ is the evolution matrix for layer $i$ (where $i\in \{1,2,\ldots,n\}$), for which there exists a relatively simple analytic solution \cite{Ohlsson:1999xb}. For both the electron number density and the nucleon densities, we have used the Standard Solar Model (SSM) \cite{Bahcall:2004pz}.

For the neutrino interaction part, both neutral-current (NC) and charged-current (CC) interactions with nucleons need to be taken into account\footnote{The cross-sections for interactions with electrons are negligible in comparison.}. In a NC interaction, there is no measurement of the neutrino flavor, and thus, the neutrino flavor composition will remain the same. However, the neutrino energy will be degraded, leading to a degradation in the high-energy neutrino yields and an increase in the low-energy yields. On the other hand, CC interactions are flavor dependent. When a neutrino interacts through a CC interaction, then the neutrino is lost and a charged lepton is produced. If the neutrino interacts as an electron or muon neutrino, then no new neutrinos will be produced, since electrons are stable and muons are stopped in the solar medium before they have had time to decay and give rise to new neutrinos. For tau neutrinos, there will be a secondary flux of neutrinos induced at CC interactions as the resulting tau will decay almost instantly. Regardless of the original annihilation channel, this secondary yield of neutrinos will have a flavor composition similar to that of the $\tau^+\tau^-$ channel (\ie, with a higher yield of tau neutrinos), which is the one that we have focused on in these proceedings. In addition, unlike the NC cross-sections, the CC cross-sections are not equal for all flavors at energies below a few hundred GeV (the tau neutrino cross-section is somewhat lower due to the effects of the tau mass). In our Monte Carlo, we account for this by first assuming that the cross-sections are the same and equal to the electron and muon neutrino cross-sections. We then randomize a point of interaction according to the solar nucleon density and the cross-section. The flavor of the interacting neutrino is then randomized according to the flavor state at this point. If the interacting neutrino is a tau neutrino, then we ignore the interaction with a probability appropriate for taking the suppression of the tau neutrino cross-section into account. Regardless of the type of interaction, we continue propagate any remaining neutrinos using the same algorithm. It is relatively easy to show that this Monte Carlo treatment is statistically equivalent to the more stringent approach using the density matrix formalism \cite{Cirelli:2005gh}, see the appendix of \Ref{Blennow:2007tw}.

For the examples presented here, unless stated otherwise, we will use the following values for the neutrino oscillation parameters
\begin{eqnarray*}
&&\theta_{12} = 33.2^\circ,\
\theta_{13} = 0,\
\theta_{23} = 45^\circ,\
\delta      = 0,\\
&&\Delta m_{21}^2 = 8.1 \cdot 10^{-5}\ {\rm eV}^2,\
\Delta m_{31}^2 = 2.2\cdot 10^{-3}\ {\rm eV}^2,
\end{eqnarray*}
in accordance with the best-fit values of \Ref{Maltoni:2004ei} and normal neutrino mass hierarchy.

\section{Results at the solar surface}

\begin{figure}
\includegraphics[width=0.45\textwidth,angle=0]{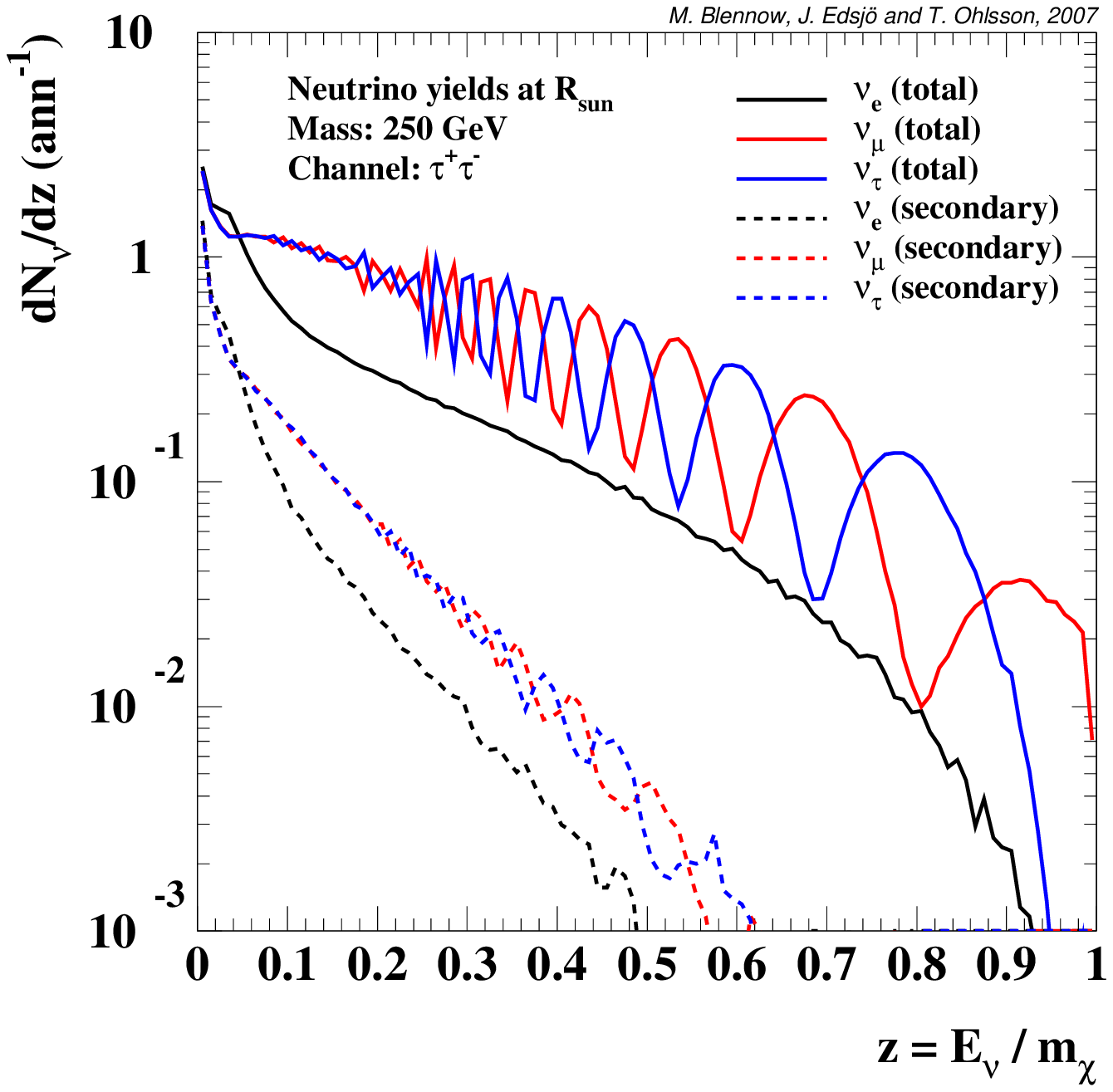}
\caption{The neutrino yields as a function of $z = E_\nu/m_\chi$ at the solar surface for annihilation of 250~GeV WIMPs into $\tau^+\tau^-$. From \Ref{Blennow:2007tw}.}
\label{fig:rsun}
\end{figure}
Figure \ref{fig:rsun} shows the total neutrino yields at the solar surface and the contribution to this yield from the regeneration due to CC tau neutrino interactions. From this figure, we can observe that the muon and tau neutrinos have mixed during the propagation out from the center of the Sun. At the same time, the electron neutrino yield is essentially unaffected by the oscillations and the main changes come from interaction effects. This is due to the facts that the Mikheyev--Smirnov--Wolfenstein (MSW) potential  is so large that the electron neutrinos essentially decouple during most of the propagation inside the Sun and that the leptonic mixing angle $\theta_{13}$ is equal to zero, and thus, no adiabatic transitions will occur at the high MSW resonance.
\begin{figure}
\includegraphics[width=0.45\textwidth,angle=0]{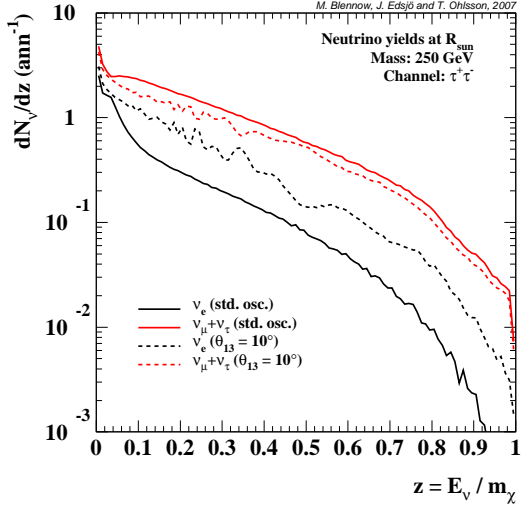}
\caption{The neutrino yields as a function of $z = E_\nu/m_\chi$ at the solar surface for annihilation of 250~GeV WIMPs into $\tau^+\tau^-$ for both $\theta_{13} = 0$ and $\theta_{13} = 10^\circ$. From \Ref{Blennow:2007tw}.}
\label{fig:rsun-t13}
\end{figure}
However, if we allow for non-zero $\theta_{13}$ as in \Fig{rsun-t13}, then there will be level transitions at the MSW resonance, which can be observed as the electron neutrino yield is significantly increased for $\theta_{13} = 10^\circ$ as compared to $\theta_{13} = 0$. Note that this effect only occurs for neutrinos in the case of normal neutrino mass hierarchy and for anti-neutrinos in the case of inverted neutrino mass hierarchy as there is no resonance for anti-neutrinos in the normal neutrino mass hierarchy and vice versa.

\section{Results at 1 AU}

\begin{figure}
\includegraphics[width=0.45\textwidth,angle=0]{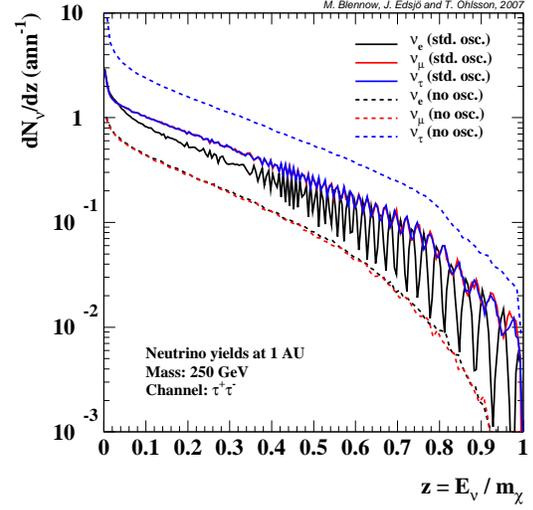}
\caption{The neutrino yields as a function of $z = E_\nu/m_\chi$ at 1~AU for annihilation of 250~GeV WIMPs into $\tau^+\tau^-$. From \Ref{Blennow:2007tw}.}
\label{fig:1AU}
\end{figure}
Once the neutrinos have been propagated to the solar surface, it is trivial to propagate them to a distance of 1~AU using the same algorithm with both the electron number density and the nucleon density set to zero. Because of this, we do not need to split the propagation into several layers of constant density as was the case of the propagation out of the Sun. The result of this propagation is shown in \Fig{1AU}, where we also show the results when neutrino oscillations are turned off. Studying this figure, we can observe that there are a few additional features compared to the result at the solar surface. First of all, the electron neutrinos are now also mixed together with the other two neutrino flavors. This mixing is due to the leptonic mixing angle $\theta_{12}$ and the small mass squared difference $\Delta m_{21}^2$ becoming important with the increasing base-line. However, as $\theta_{12}$ is not maximal, the mixing is not as complete as the one we observed between muon and tau neutrinos at the solar surface. Second, the oscillation phase in the transitions between muon and tau neutrinos can no longer be observed. The reason for this is that the neutrino oscillation phase is proportional to $L/E_\nu$, where $L$ is the neutrino base-line. Thus, as the base-line is increased from $R_\odot$ to 1~AU, the oscillation phase varies more rapidly with energy. As a result, the oscillation phase can no longer be observed at the binning provided in the figure (and also not in reality due to the finite energy resolution of detectors).

Finally, although not visible in the figure, the effects of having non-zero $\theta_{13}$ are similar to those for the results at the solar surface. However, because of the fact that there is also mixing due to $\theta_{12}$ and $\Delta m_{21}^2$, the difference between the $\theta_{13} = 0$ and $\theta_{13} = 10^\circ$ scenarios is not as pronounced.

\section{Results at an actual detector}

For any real detector located at the Earth, the distance to the Sun will vary throughout the year. In addition, depending on the specific time, neutrinos may need to travel longer or shorter distances through the Earth in order to reach the detector. In the final step of our Monte Carlo, we have the possibility to take the neutrino states at 1~AU and propagate them to the specific detector location.\footnote{Note that if the Earth is closer to the Sun than 1~AU, then we first need to apply inverse propagation to find the neutrino states that enter the Earth.} If we want to have a time-average at the detector, then we can also randomize the time of the neutrino event and compute the detector location from this time. For this we use a somewhat simplified model for the Earth orbit where both perihelion and the winter solstice occur at New Year (both are off by about one week but the effect is small). Future releases of the Monte Carlo code may include a more exact astronomical model.

\begin{figure}
\includegraphics[width=0.45\textwidth,angle=0]{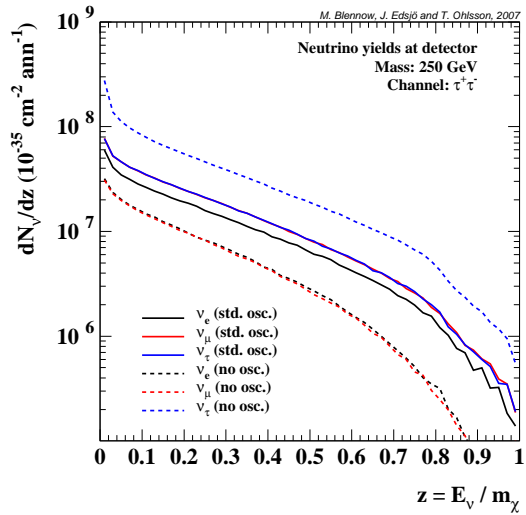}
\caption{The time-averaged neutrino yields as a function of $z = E_\nu/m_\chi$ at the South Pole for annihilation of 250~GeV WIMPs into $\tau^+\tau^-$. From \Ref{Blennow:2007tw}.}
\label{fig:detector}
\end{figure}

In \Fig{detector}, we show the resulting time-averaged neutrino yields for a detector at a latitude of $-90^\circ$ (\ie, the South Pole) with a viewing period of half a year (vernal to autumnal equinox).
As can be observed, the main effect of the propagation to the detector is that the oscillatory pattern has been smoothened out. The reason for this is mainly that the time-averaging is an effective average of the base-line. Thus, the oscillation phases will average out due to their $L/E_\nu$ behavior. In principle, the oscillatory pattern should remain if the statistics for smaller parts of the year were large enough or if we instead observed the $L/E_\nu$ behavior. In reality, the energy resolution of any realistic neutrino telescope will not be good enough to resolve this. However, the effect could be observed in some particular WIMP models where the WIMPs have a large annihilation branching ratio directly into neutrinos. For such WIMP models, we would be provided with monochromatic neutrinos and the energy resolution of the neutrino telescope would become irrelevant.\footnote{Note that, in order to observe neutrino oscillations in such a model, there would also need to be an asymmetry in the initial fluxes of different neutrino flavors.}
Another interesting observation is that, while the final electron and muon neutrino yields are equal when neutrino oscillations are turned off, the muon and tau neutrino yields are equal when neutrino oscillations are turned on.

\section{Summary and conclusions}

We have seen that neutrino oscillations can result in significant changes in the neutrino spectra from WIMP annihilations in the Sun. The neutrino yield from annihilations of 250~GeV WIMPs into $\tau^+\tau^-$ has been used in order to illustrate this fact. Our results have shown that muon and tau neutrinos generally mix with each other already during the propagation out of the Sun, while the electron neutrinos are mainly mixed during the vacuum propagation to the Earth.

Our results are also relatively insensitive to the exact neutrino oscillation parameters. The leptonic mixing angles mainly affect the degree of mixing, while the actual value of the mass squared differences is more or less irrelevant as long as they stay within the order of magnitude indicated by neutrino experiments. This is due to the fact that the oscillation phases will be averaged out in any reasonable experiment.

Since neutrino telescopes will be mainly sensitive to muon neutrinos, we should also think about the possible effects of neutrino oscillations on the prospects for the detection of DM. For neutralino DM, one usually expects to produce less tau neutrinos. Hence, in such models, the muon neutrino flux at the detector is generally decreased by neutrino oscillations, since more muon neutrinos will oscillate into tau neutrinos than vice versa. However, for Kaluza--Klein DM, there can be a significant branching ratio into charged leptons. Although the branching ratios to charged leptons will be equal for all flavors (neglecting the effect of the lepton masses), only the annihilations into $\tau^+\tau^-$ will give rise to high-energy neutrinos detectable at neutrino telescopes. As this annihilation channel has a significant overproduction of tau neutrinos, which can subsequently oscillate into muon neutrinos, the muon neutrino flux at the detector can receive a large increase (about a factor of four) compared to the non-oscillatory case.

\end{document}